\documentclass{emulateapj}
\usepackage{natbib} 
\usepackage{url}
\begin{document}
\title{Creating updated, scientifically-calibrated mosaic images for the RC3 Catalogue}
\author{Jung Lin Lee\altaffilmark{1}, 
Robert J.~Brunner\altaffilmark{2,3,4,5},
}
\altaffiltext{1}{Department of Astronomy, University of California Berkeley, Berkeley, CA 94720; \ {\it dorislee@berkeley.edu}}
\altaffiltext{2}{Department of Astronomy, University of Illinois, Urbana, IL 61801;\ {\it bigdog@illinois.edu}}
\altaffiltext{3}{Department of Statistics, University of Illinois, Champaign, IL 61820 USA}
\altaffiltext{4}{National Center for Supercomputing Applications, Urbana, IL 61801 USA }
\altaffiltext{5}{Beckman Institute for Advanced Science and Technology, Urbana, IL 61801 USA}

\begin{abstract}
The Third Reference Catalogue of Bright Galaxies (RC3) is a reasonably complete listing of 23,011 nearby, large, bright galaxies. By using the final imaging data release from the Sloan Digital Sky Survey, we generate scientifically-calibrated FITS mosaics by using the montage program for all SDSS imaging bands for all RC3 galaxies that lie within the survey footprint. We further combine the SDSS $g$, $r$, and $i$ band FITS mosaics for these galaxies to create color-composite images by using the STIFF program. We generalized this software framework to make FITS mosaics and color-composite images for an arbitrary catalog and imaging data set. Due to positional inaccuracies inherent in the RC3 catalog, we employ a recursive algorithm in our mosaicking pipeline that first determines the correct location for each galaxy, and subsequently applies the mosaicking procedure.
As an additional test of this new software pipeline and to obtain mosaic images of a larger sample of RC3 galaxies, we also applied this pipeline to  photographic data taken by the Second Palomar Observatory Sky Survey with $B_J$, $R_F$, and $I_N$ plates. We publicly release all generated data, accessible via a web search form, and the software pipeline to enable others to make  galaxy mosaics by using other catalogs or surveys.
\end{abstract}

\section{Introduction}

Astronomers have a long history of cataloguing objects for subsequent study, for example the Messier catalog~\citep{messcat} and the New General Catalog~\citep[NGC;][]{ngccat} have provided valuable guidance to help astronomers study objects with similar properties. Today, we have entered the era of big data where large surveys such as the Sloan Digital Sky Survey~\citep[SDSS;][]{sdsstech} have uniformly surveyed large fractions of the entire sky, providing detailed photometric and astrometric information for millions of sources. However, the software pipelines that  process these valuable data are optimized for the more numerous small sources. As a result, large, nearby, and bright galaxies are essentially treated as contaminants. Yet these galaxies remain incredibly important: providing detailed insight into the dynamics of galaxies and serving as a low redshift sample with which we can compare higher redshift galaxies against for better understanding of the evolution of these galaxies. 

One of the more popular catalogs of nearby galaxies is the Third Reference Catalog of Bright Galaxies~\citep[RC3;][]{rc31991}, which contains 23,011 galaxies with an apparent diameter greater than one arcminute at the $D_{25}$ isophotal level, with a total $B$-band magnitudes brighter than $~15.5$, and with a redshift not in excess of 15,000 km/s. The overall catalog is supplemented by selected galaxies that may only meet one or two of these conditions as well as some  nearby compact galaxies. Given the efficacy of this catalog, previous authors have used this sample as the basis for making image mosaics for a large sample of galaxies. 

The first such effort was made by~\cite{hbrc3}, who made color-composite images of selected RC3 galaxies by using the SDSS $g$, $r$, and $i$ band images from the sixth SDSS data release. A subsequent effort by \citet{efigi}, dedicated to the study of galaxy morphology, generated a set of 4,458 FITS and color images by using the SDSS DR4 data. This latter effort employed a visual inspection to remove artifacts and galaxies with missing data. Finally, a separate effort, known as the NASA-Sloan Atlas\footnote{\url{http://www.nsatlas.org}}, has been undertaken to construct and analyze a complete set of galaxies within approximately 200 Megaparsecs~\citep{nsa}, which is the approximate redshift limit of the RC3 catalog.

Despite the diverse set of studies that incorporates the RC3 catalog mentioned earlier, there is no dedicated study that uniformly mosaic all of the RC3 sources with updated astrometry. As the SDSS project has published their last imaging data release~\citep[DR10; ][]{dr10}, which includes the final photometric and astrometric calibrations, we have decided to revamp these previous efforts to make scientifically-calibrated image mosaics and updated positional coordinates for all RC3 galaxies that lie within the SDSS DR10 footprint. 

	Using a source update algorithm, we can recursively identify and mosaic the central RC3 galaxy and, in complement, use the mosaic outputs to find large extended RC3 sources. We further generalize this software framework  for use on any arbitrary user-defined catalog using imaging data from any scientifically-calibrated sky survey . This new pipeline is also applied to digitized photgraphic plate data from  Second Palomar Observatory Sky Survey~\citep[POSS-II;][]{poss2}. The complete sky coverage of the POSS-II survey enables us to generate updated coordinate positions for all the RC3 galaxies with inaccurate catalog coordinates.  Our pipeline can be easily applied to existing data such as the Two Micron All Sky Survey~\citep[2MASS;][]{2mass}, or to future surveys such as the Dark Energy Survey~\citep[DES;][]{des} or the Large Synoptic Survey Telescope~\citep[LSST;][]{lsst}. 
The scientifically calibrated FITS image mosaics generated by the pipeline for a specific galaxy catalog can be used both for individual source studies, as well as ensemble studies of source populations. In addition, these mosaics can be combined with calibrated mosaic images at different wavelengths into new, multi-band color images by using mosaic images from other surveys. This can be especially beneficial when a survey has only one or two bands, such as the POSS-I~\citep{poss1} or GALEX~\citep{galex}, as long as the image sizes are adjusted to match. 

The SDSS and POSS-II mosaic images that we have generated and publicly released may be useful for commissioning and data processing of the next-generation surveys such as the DES and LSST. The updated RC3 coordinates indicate regions of the sky that may be affected by these large galaxies, and the FITS mosaic images can be used to model-fit the galaxy's shape and light distribution. Alternatively, this same information can be used to place spectroscopic fibers on RC3 catalog sources. 

In Section~\ref{data-sec}, we introduce the RC3 catalog and the SDSS and POSS-II data that we use to generate mosaic images. Section~\ref{mosaic-sec} introduces our software pipeline and the relevant algorithmic details that enable us to obtain improved astrometric precision for the catalog galaxies. We discuss the pipeline performance and evaluate the science-quality for the pipeline results for the SDSS and POSS-II data in Section~\ref{results-sec}, before concluding the paper and discussing the overall project in Section~\ref{conc-sec}. 

\section{Data}\label{data-sec}

To construct large, calibrated image mosaics, we need a catalog of galaxies and an image data set. A number of different candidate galaxy catalogs exist; we also could follow the example used to develop the NASA-Sloan Atlas and construct a new catalog based on specific physical criteria. We choose, however, to use the RC3 catalog of galaxies~\citep{rc3}, which we detail in \S~\ref{sec:rc3}. For our imaging data, we actually selected two different surveys: the SDSS and the POSS-II. The SDSS uniformly surveyed a large fraction of the sky with digital detectors, while the POSS-II is an older photographic plate survey that covers nearly the entire sky. We discuss these two imaging surveys in detail at the end of this section. While not directly discussed in this paper, our pipeline approach could easily be extended to other datasets with archived, calibrated image data products, such as the Two Micron Sky Survey~\citep{2mass}, the Wide-field Infrared Survey Explorer~\citep{wise}, and the Galaxy Evolution Explorer~\citep{galex}.

\subsection{The RC3 Catalog\label{sec:rc3}}

The simplest approach to construct a uniform galaxy catalog is to  define a sample by limiting apparent brightness within a survey. Original attempts to accomplish this task date back to the original Harvard Survey of External Galaxies~\citep{shapley-ames}, which contained 1,249 objects brighter than 13$^{th}$ magnitude. Many of these galaxies were subsequently included in the \textit{University of Texas, Monographs in Astronomy}, which were predecessors to the RC3 catalog. The actual RC3 Catalog is an update to the Original and Second Reference Catalog of Bright Galaxies~\citep{rc2}. The original RC3 galaxy catalog compiled by \citet{rc3}, contains a  complete listing of 23,011 galaxies with $D_{25}$ apparent major isophotal diameter greater than one arcminute and with a total B-band magnitude greater than 15.5$^{th}$ magnitude. 

An update of the RC3 catalog was published a few years after the original RC3 catalog by~\citet{rc3-94}. In this project, we use the RC3 catalogue information available through the VizieR Service\footnote{\url{http://vizier.u-strasbg.fr/viz-bin/VizieR?-source=VII/155}}as our starting conditions,  which contains the updated RC3 data published in 1994. As the imaging data used to update these galaxies  came from various different imaging programs, the final, updated catalog is heterogeneous with a non-uniform distribution of updated galaxies. The RC3 data from  NASA Extragalactic Database (NED) were last updated from the 1993 version 3.9b of the RC3 catalog. Until 2011, it was incrementally updated with new, published observations and subsequently maintained by Harold G. Corwin\footnote{\url{http://haroldcorwin.net/rc3/bugs.rc3}}. However, only the few RC3 sources that happen to overlap with other newer source catalogs are updated. \footnote{For example, the astrometry of PGC 2557 was last updated in 2010 with the Chandra source catalog ~\citep{chandra}.} In this project, we uniformly update the coordinate positions  of all RC3 sources that lie in the catalog.

Since the RC3 catalog  is a reasonably complete representation of large, bright nearby galaxies in the extragalactic sky, it remains a popular catalog. Selected galaxies or complete subsets have been used in astrophysical studies of quasars and X-ray sources~\citep[e.g.,][]{walton-rc3}, and for galaxy morphology and clustering studies~\citep[e.g.,][]{best-rc3, knapen-rc3}.  The RC3 catalog also serves as a basis for statistical studies  in cosmology, for example within the New York University-Value Added Galaxy Catalog~\citep{nyuvagc}. However, despite different levels of precision due to the non-uniform astrometric updates, many survey catalogs and studies have still taken the NED RC3 positions to be the de-facto positions of the RC3. In this project, we present a complete updated set of all the RC3 sources more accurate than the NED coordinates that could be used for such future studies. 
\subsubsection{Updating the RC3 Astrometry\label{sec:position}}

As described in \S\ref{mosaic-sec}, we use the Montage toolkit to reproject and mosaic astrometric and photometric calibrated FITS images. Initially when we followed the standard mosaicking steps in Montage, we obtained many mosaics with off-centered or missing RC3 galaxies. After further review, we realized that this effect was due to the inherent positional inaccuracy in the catalog, which, given the size of a typical RC3 galaxy, was surprising. On further reflection, however, this effect can be understood by the heterogeneous origin and updating of the RC3 catalog.

The entries in the RC3 catalog hosted by VizieR are astrometrically tied to the B2000.0 FK4 reference system. Due to the heterogenous positional updates, the RC3 galaxies are denoted with two different levels of accuracy: HH MM SS.s, DD MM SS for positions that have been updated  to an accuracy of approximately $5$--$8$ arcseconds, and  HH MM.m, DD MM for galaxies whose positional accuracy remains at approximately $1$--$2$ arcminutes as presented in the original catalog. The 1991 version of the RC3 catalog~\citep{rc31991} contains 5,492  galaxies that fall in the latter group. As a result, we developed an iterative mosaicking step within our pipeline, as discussed in~\S\ref{pos-sec}, to compute updated astrometric coordinates for those galaxies with poor astrometry so that we would have the target galaxy centered on the final mosaic.

\begin{table}
\footnotesize
    \begin{tabular}{llll}
    \hline
    ~                           & SDSS    & POSS-II             \\ \hline
    Imaging bands               & u, g, r, i, z  & $R_F$, $B_J$, $I_N$             \\
    Sky Coverage (\%)                & 35.28 & 78.27             \\
    Resolution ($^{\prime\prime}$/pix) & 0.396  & 1.7                 \\
    Imaging Technique           & CCDs    & Photographic Plates\\ \hline
    \end{tabular}
    \label{table:comptbl}
    \caption{{\footnotesize A summary of the two surveys processed in this work.}}
\end{table}

\subsection{SDSS}
The SDSS imaging data is acquired from the $2.5$-m telescope~\citep{sdss-tel} at the Apache Point Observatory in New Mexico. For our mosaics, we use the imaging data from the SDSS Data Release 10. The SDSS imaging camera~\citep{sdss-camera} was one of the first, large imaging cameras that leveraged Charge Coupled Devices (CCDs) as photon recording instruments. CCD detectors have obvious throughput advantages over photographic plates and also provide a near-linear response to input signals. In addition, the pixel resolution and mean seeing at Apache Point provide a reasonable angular resolution, especially for large, bright galaxies. 

The SDSS imaging data is processed by the $\texttt{photo}$ pipeline~\citep{sdss-photo}, which is responsible for a number of different image processing procedures, including source detection, deblending, model-fitting, and astrometric and photometric calibration. We quantify the imaging quality of the SDSS data using the $\texttt{clean}$ flag\footnote{\url{http://skyserver.sdss3.org/dr10/en/help/cooking/general\\/flags6.aspx}}, which is extracted from the SDSS SkyServer. For extended sources like a galaxy, the $\texttt{clean}$ flag is defined from data masks with variables that quantify imaging quality metrics such as the PSF magnitude error, cosmic rays, or undefined profiles from model fits. We extract parameter information by querying the SDSS SkyServer via the SDSS Command Line Query Tool, which retrieves the calibrated, sky-subtracted corrected ($\texttt{fpC}$) frames with the calibration meta-data in bulk from the Science Archive Server.

\subsection{POSS-II\label{POSSII}}
The Second Palomar Observatory Sky Survey (POSS-II) was a photographic survey that covered most of the night sky, including the northern region later covered by the SDSS. POSS-II was an improved update to the original National Geographic Society-Palomar Observatory Sky Survey~\citep[NGS-POSS or POSS-I;][]{ngs-poss}, which was one of the first, large area photographic surveys. We acquire FITS images for all three plate types: $B_J$, $R_F$, and $I_N$ as required to construct mosaic images from the Space Telescope Science Institute's Digitized Sky Survey project~\citep[DSS;][]{dss}.  These photographic plates were calibrated to the  Gunn $g$, $r$, $i$ bands by a separate CCD observing campaign, producing the Digitized Palomar Sky Survey data~\citep[DPOSS;][]{dposs}. 

Our software pipeline obtains photographic plate data from POSS-I, POSS-II and UK Schmidt Telescope Survey by using the NASA/IPAC Infrared Science Archive's (IRSA) Finder Program Interface. \footnote{\url{http://irsa.ipac.caltech.edu/applications/FinderChart/docs\\/finderProgramInterface.html}} Each photographic plate covers $6.5^{\circ} \times 6.5^{\circ}$ of the sky with centers spaced approximately $5^{\circ}$ apart (thus providing significant overlap between adjacent plates). After lossy compression, each plate is approximately 1.1 GB, although a crop-out of the plate can be retrieved to reduce the download time. Due to the large size of these plates, our pipeline rarely needs to stitch multiple, adjacent fields; instead we simply use the full, single-plate image data to complete the positional update algorithm.

Even though POSS-II has the advantage of a greater sky coverage than SDSS, it suffers from several disadvantages that do impact our photographic-plate based mosaics. First, photographic plates are much less sensitive to incoming photons ($\sim$1\% quantum efficiency) as compared to CCDs ($\sim$80\%), thus the photographic-based mosaics have a reduced clarity. Second, objects detected near the edges of photographic plates can suffer from vignetting, which may result in inaccurate astrometric calibrations being applied to those objects. Finally, photographic plates have a non-linear response to incoming photons, thus the absolute photometric calibration of photographic plates is less accurate, which will affect multi-plate mosaics. However, since our pipeline does not perform additional photometric calibration, this factor does not affect the output mosaics.

\section{Mosaic Pipeline and algorithms}\label{mosaic-sec}
\begin{figure}[h]
\includegraphics[width=0.5\textwidth]{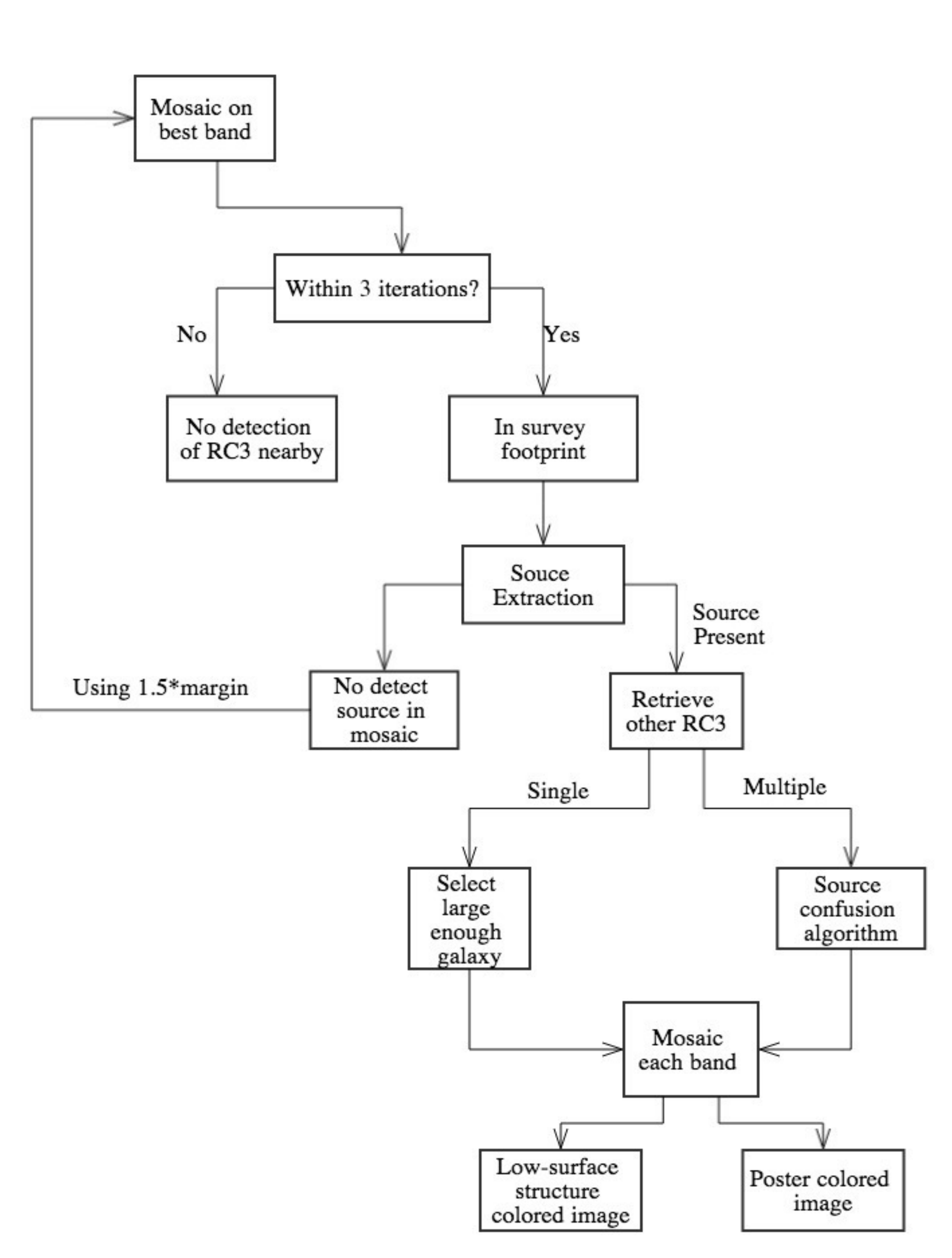}
\caption{A flowchart detailing the basic steps employed by our pipeline to first construct calibrated FITS image mosaics for each photometric band in an imaging survey of the target galaxy, followed by the construction of color-composite images for the target galaxy.}
\label{flowchart}
\end{figure}

Given the size and positional inaccuracy of the RC3 catalog, we decided to construct a pipeline to automate the construction of the FITS mosaic images and the color-composite images for each RC3 galaxy contained within a given imaging survey like the SDSS. This pipeline is presented as a flowchart in Figure~\ref{flowchart}, and is implemented in the Python programming language.

Broadly speaking, our pipeline leverages several open-source tools: Montage~\citep{montage}, Astropy~\citep{astropy}, and  STIFF~\citep{stiff}. Montage is used to reproject each input image appropriately, and to combine the reprojected images into the final calibrated mosaic FITS image. STIFF, on the other hand, is used to make the color-composite images from the mosaicked FITS images generated by montage. While Montage can make color-composite images, we found that STIFF was easier to use within our pipelined approach since STIFF provides more flexible parameters for the optimal color-mapping.

In the rest of this section, we first detail several new software tools we developed to generate more accurate astrometry for mosaicing each RC3 galaxy. Next, we provide a detailed discussion of  specific software components, including object classes that generalizes our approach to support different galaxy catalogs and different photometric surveys. We present a flowchart that highlights the operational steps of our pipeline in Figure~\ref{flowchart}.

\subsection{Algorithm Enhancements}

The Montage software can be accessed via the Virtual Astrophysical Observatory (VAO) Image Mosaic Service at the IRSA website to robustly generate mosaic images from multiple surveys, including from both the SDSS and DSS. Given the nature of the RC3 catalog, however, we had to overcome two challenges when including Montage as part of our overall mosaicking pipeline. First, the relatively large positional uncertainty present for some galaxies within the RC3 catalog meant a mosaic could be automatically generated can either have the target galaxy off-centered, or worse entirely missing from the mosaic. Second, since galaxies are often in groups or clusters, a positional uncertainty can lead to a mis-identification of the target galaxy within the constructed mosaic. As a result, two components are developed within our pipeline to address these issues in order to generate appropriately-centered mosaic images.

\subsubsection{Positional Update\label{pos-sec}} 

We created a class, called $\texttt{RC3}$, to encapsulate the data and methods relevant to the RC3 sources. Each $\texttt{RC3}$ instance contains the catalog RC3 coordinate information, any updated coordinate information, radius, and a unique identifier. While we could have adopted a standard numerical identifier scheme, we instead chose to adopt each galaxy's Catalogue of Principal Galaxies (PGC) number, since this is a unique identifier for each RC3 galaxy that is present in the original RC3 catalog. This unique name is used for naming the generated data products accessible on our website.

To overcome the problem of inaccurate positions discussed in \S\ref{sec:position}, we first find the target galaxy in the imaging data and then use this position to update the astrometric information appropriately. This algorithm is implemented in the $\texttt{source\_info}$ method of the $\texttt{RC3}$ class, and first generates a mosaic with a field of view roughly six times larger than the radius of the target galaxy.  Next, SExtractor is used to detect all sources on the newly-generated mosaic. The ratio of RC3 to mosaic size is chosen to ensure that SExtractor can  determine an accurate estimate of the background sky level for conducting its measurements. Of all detected sources, only those with a radius greater than 5.94 arcseconds are retained. This size was empirically selected to eliminate most stellar sources and background noise spikes while still retaining the subset of RC2 galaxies contained in the RC3 catalog that are smaller than one arcminute as described by~\citealp{rc2}. Montage has a built-in set of modules that rectifies the background and subtracts the overlapping, neighboring fields in the raw input image tiles by least square fitting solution ~\citep{montage} . Since the sole purpose of these initial $r$-band mosaics is to improve astrometry, we do not rectify the background in these mosaics in order to speed the computation by minimizing data I/O.

At this point, our pipeline is in one of three states depending on whether the new mosaic contains zero, one, or more RC3 candidates detected by SExtractor. First, if there are zero candidate RC3 galaxies in the new mosaic, we recursively create mosaics with a field that is 150\% larger in size than the original until either a RC3 candidate galaxy is detected or we exit after three iterations and declare the RC3 source as not found. If only one candidate galaxy is detected, the pipeline proceeds directly to the final image generation step. This ratio was chosen so that if multiple large galaxies are detected, the pipeline executes the source confusion algorithm as outlined in \S\ref{sec:sc}. At the end of this process if at least one candidate RC3 galaxy is detected, then we generate mosaic images in all bands covered by the survey with Montage's background rectification procedures enabled. In our analysis in \S\ref{photo}, we find that the background rectification modules in Montage is necessary for preserving the photometric quality of the science-grade image by removing a constant magnitude offset of around +1.5 in the output mosaic. 

Once the mosaic images have been constructed in all bands for the given survey, we select images observed through three different filters and recombine them into a color mosaic image by using STIFF\label{sec:best_low}. The three bands selected are the $g$, $r$, and $i$ bands for the SDSS, which has five bands, and the $B_J$, $R_F$, and $I_N$ bands for the DSS survey, which only has these three bands. At this point, two color images are generated (although this number can be altered if necessary). The first color image generated emphasize low surface brightness structures, such as the halo around a galaxy, or interacting tidal streams. The second color image generated is a poster or publication quality image that uses higher background cuts to ensure a clean, highly contrasted image. We note that to construct a color image, our pipeline does require at least three bands. Fortunately, most photometric surveys meet this criterion, which is not surprising since the construction of a color-color plot, used to minimize the effects of extinction and identify stellar or extragalactic source populations, require at least three bands~\citep[see, e.g.,][]{2mass}. 

Our pipeline is more than just a wrapper around Montage and STIFF, as it is necessary for successfully mosaicing these heterogeneous catalog sources. Our software tool is distinguished by its unique relationship in how the positional update algorithm informs the mosaicking process and vice versa. Without the updating algorithm, a standard mosiacking procedure will not be able to find and center the mosaic around the source of interest and will instead simply mosaic an empty piece of the sky. 

\subsubsection{Source Confusion\label{sec:sc}}

\begin{figure}
\centering
  \includegraphics[width=.225\textwidth]{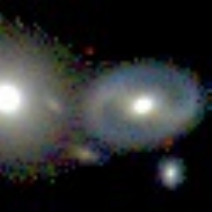}
  \includegraphics[width=.225\textwidth]{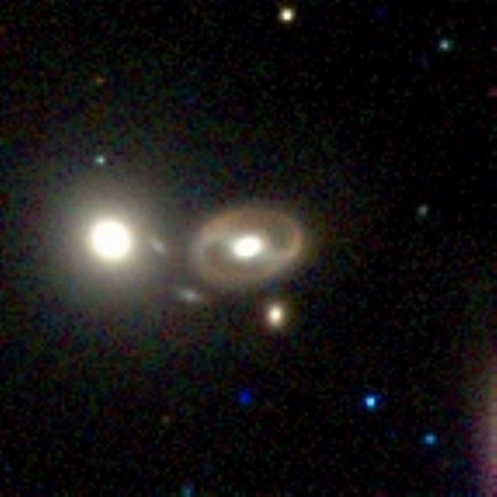}
\caption{ A visual demonstration of our source confusion algorithm using SDSS data, see Figure~\ref{fig:positional_update_plot} for a wider angle view of this same field. (Left) A color-composite image made from mosaicked $g$, $r$, and $i$ FITS images. (Right) The regenerated color-composite image made from the original data after the source confusion algorithm has been applied, resulting in a recentering on the target galaxy PGC58.}
\label{fig:SCdemo}
\end{figure}

\begin{figure}[h]
\includegraphics[width=0.5\textwidth]{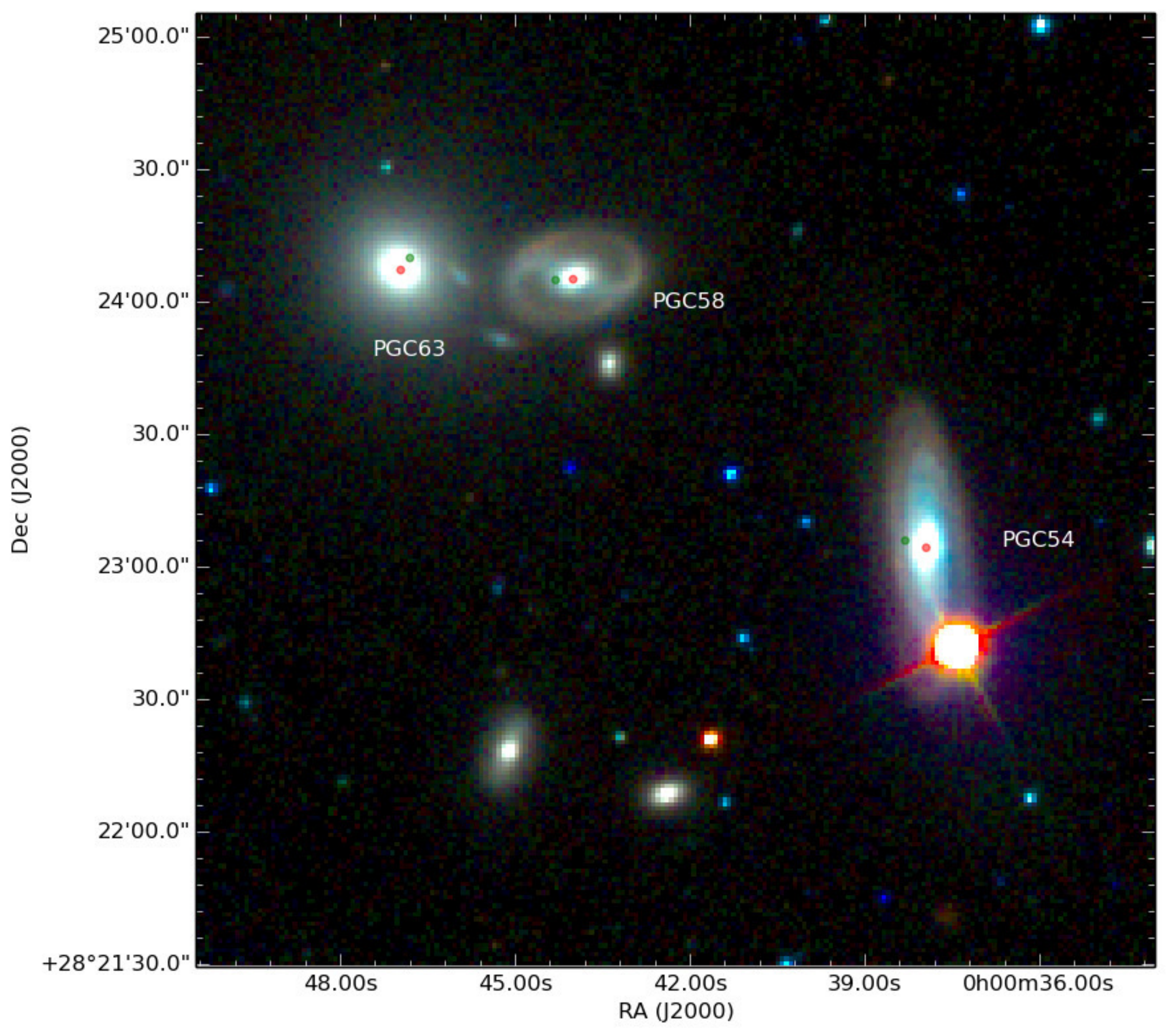}
\caption{A visual demonstration of the results of the source confusion algorithm for three RC3 Galaxies showing the original RC3 catalog galaxy coordinates (green markers) and our newly calculated coordinates (red markers).}
\label{fig:positional_update_plot}
\end{figure}

The second challenge our pipeline was forced to tackle was the source confusion problem. Since large galaxies are often physically located near other galaxies, any astrometric error in the location of our target RC3 galaxy could result in source confusion on a generated mosaic (see Figure~\ref{fig:SCdemo} and \ref{fig:positional_update_plot} for a visual demonstration). To tackle this challenge, we first assumed that any galaxy large enough to cause confusion would also be present in the original RC3 catalog. This assumption is supported by the fact that the RC3 catalog is reasonably complete for galaxies with apparent diameters larger than one arcminute. 

The algorithm we developed to overcome this challenge first identifies all RC3 galaxies that may lie within the newly generated mosaic image, and second matches this list to the list of galaxies that were actually detected by SExtractor within the mosaic image. The list of RC3 galaxies is generated by the $\texttt{otherRC3}$ method that is part of the $\texttt{Server}$ class described in \S~\ref{sec:server}. By default, this method queries the VizieR Catalog to obtain this information. Alternatively, this method can be overridden if a survey provides this information directly, like the SDSS, which contains an RC3 table in its SkyServer database. 

Naively, this cross-match process might appear to be simple; however, since each of these RC3 galaxies may have inaccurate astrometry, the problem becomes non-trivial. To tackle this challenge, we make the reasonable assumption that any positional inaccuracies are due to instrumental or measurement errors that might affect a galaxy's absolute position but not the relative positions of each galaxy. Thus we identify the \textit{n} RC3 galaxies that potentially lie within the mosaic image from VizieR, and compute all possible relative positional differences between these candidate galaxies. Next, we compare this list with the positional differences computed between the galaxies detected in the actual image by SExtractor. From this list of cross-matched differences, we identify each RC3 galaxy in the image, correct the overall astrometry for each detected RC3 galaxy, and regenerate a new mosaic centered on the correct location). We have verified the validity of these assumptions, as our algorithm demonstrated a success rate of 99.97\% compared with the results from the SDSS Skyserver database. Furthermore, we have tested the robustness of this approach to correctly resolve up to five RC3 sources in a single mosaic and found that none of the fields contain more source-confused RC3 beyond that. 
	
\subsection{Pipeline}

\begin{figure}[h]
\includegraphics[width=0.5\textwidth]{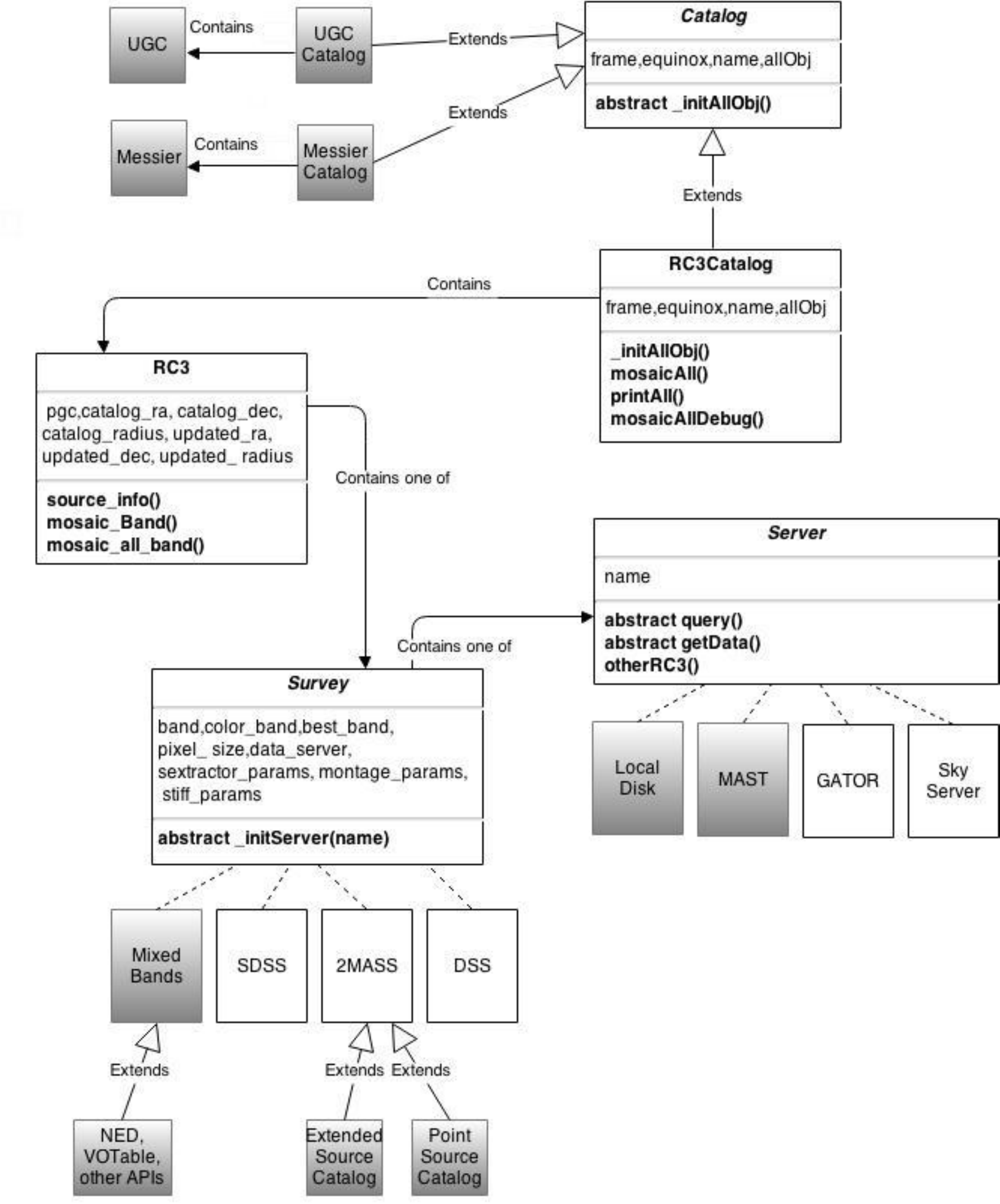}
\caption{A unified modeling language (UML) class diagram showing the relationships between different classes in the mosaicking pipeline. Grey-filled boxes shows possible extensions of this pipeline.
\label{fig:hierarchy}
}
\end{figure}

To simplify the application of our approach to other catalogs and/or surveys, we developed an automated pipeline to generate the FITS image mosaics and color-composite images. This pipeline is built by using classes that encapsulate specific data that might be relative to a certain catalog, like the RC3, or to a particular survey, like the SDSS. Figure~\ref{fig:hierarchy} highlights the overall class relationships and three abstract classes: $\texttt{Survey}$, $\texttt{Catalog}$, and $\texttt{Server}$, which simplify the incorporation of new catalog data or surveys into our pipeline. In the following subsections, we discuss several of these classes in more detail.

\subsubsection{Survey \label{sec:survey}}
The \texttt{Survey} class provides a convenient way of extending the mosaicking pipeline to take input data from another sky survey. Additional  \texttt{Survey} classes enables a user to obtain multi-band or higher resolution images for the same source object in a catalog.  In addition, extending the pipeline to an all-sky survey (such as POSS-II) increases the coverage of sources in the catalog, which leads to a more complete set of updated astrometry.  As shown in \S\ref{photo}, our pipeline preserves the photometry of the  input images.  Therefore, in order to get science-grade, calibrated output mosaics, the survey's input images must be photometrically-calibrated. In other words, our pipeline does not perform any additional calibration to the raw images. As this pipeline is intended for use on large, general, sky surveys, the task of calibrating images and tuning telescope-specific parameters are abstracted to the dedicated photometric pipeline used to generated the data products of most sky surveys. 
\subsubsection{Server\label{sec:server}}

The primary abstract class used by the pipeline is the $\texttt{Server}$ class, which encapsulates the two main tasks of data acquisition: querying imaging data and retrieving the imaging data from the actual server. Many recent surveys provide an application programming interface (API) that enables data access either by using SQL or a customized query mechanism. To actually implement a subclass of the $\texttt{Server}$ class, a mapping between the positional values of galaxies in a particular catalog and the recorded image data for the particular survey must be established. For example, SDSS image frames are uniquely identified by a particular combination of \textit{run}, \textit{camcol}, and \textit{field}, while 2MASS identifies images with a sexagesimal, equatorial position-based source name. For those surveys where this type of mapping has not already been established, such as the imaging data from the POSS-II survey, a subclass can establish a new naming scheme.

In addition to these primary tasks, each $\texttt{Server}$ subclass must also implement functionality to build and execute queries as required by the Montage mosaic software. By using a $\texttt{Server}$ class as opposed to a $\texttt{DataObject}$ class, we enable code reusability across various surveys that can be accessed via common server tools. For example, this approach allows the pipeline to easily use the IRSA GATOR query service~\citep{irsa} and Astroquery. \footnote{\url{http://astroquery.readthedocs.org/}}

\subsubsection{Catalog Objects}

\begin{figure}[h]
\centering
\includegraphics[width=0.4\textwidth]{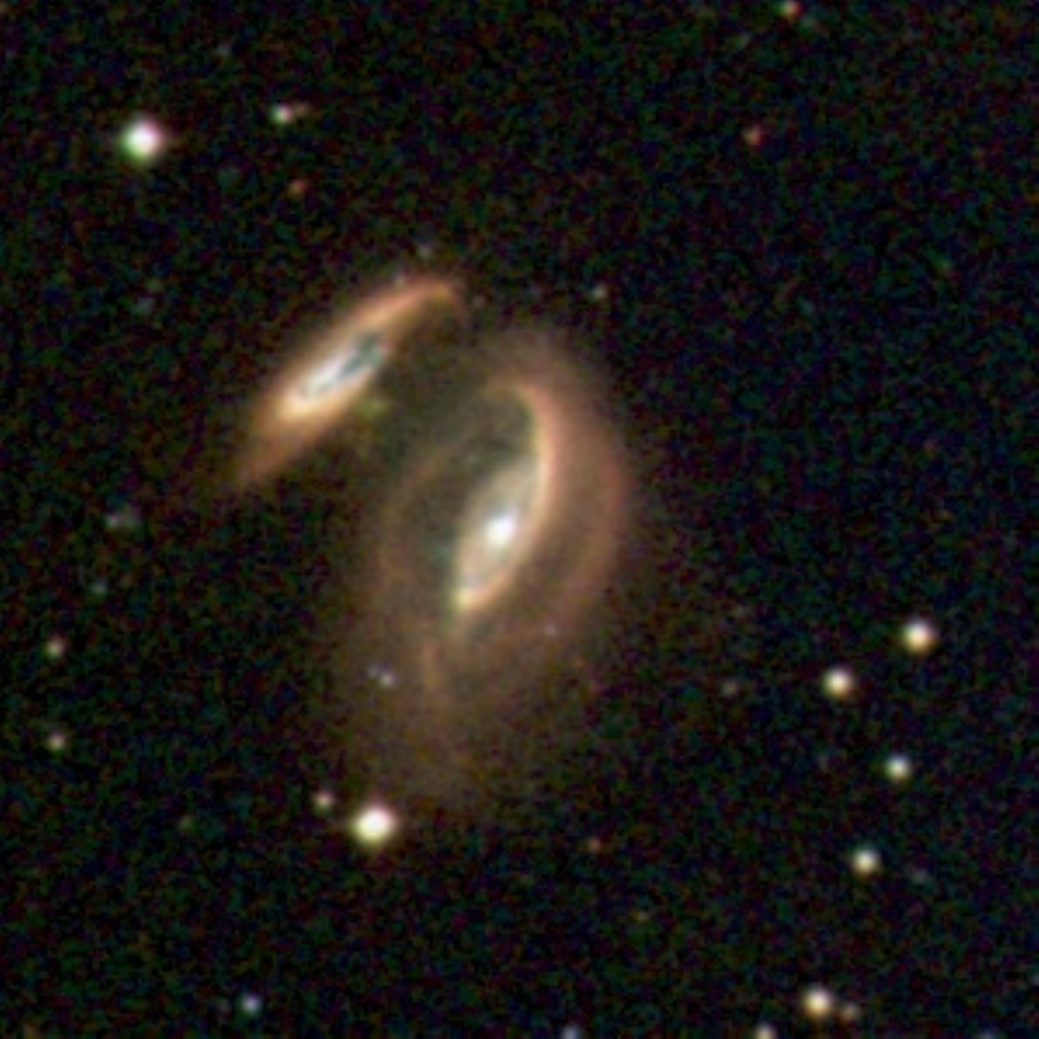}
\includegraphics[width=0.4\textwidth]{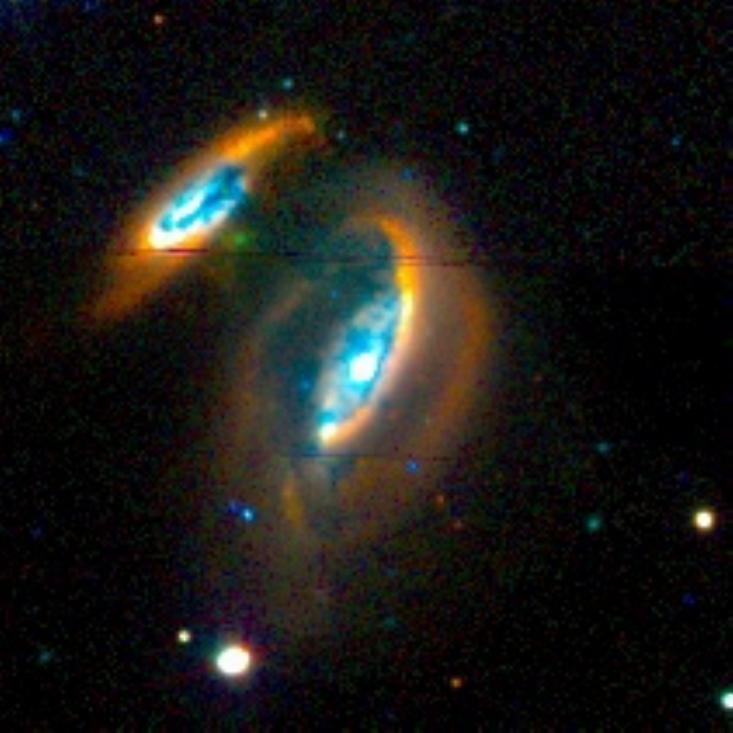}	
\caption{Sample color mosaics of PGC120 mosaic from POSS-II (top) and SDSS (bottom). The two images are of different scale.}
\label{fig:comparison}
\end{figure}

The current version of the pipeline, which solely generates mosaic images of RC3 galaxies, does not contain an abstract $\texttt{CatalogObject}$ class. As a future extension, this might be a beneficial addition as it would be helpful to have a generic class that provides a similar functionality as the $\texttt{RC3Objects}$ class. These functions contain basic information about the particular object and are survey-independent. They also perform the essential mosaicking features on a per-object basis. Therefore, these functions not only can be used for comparing resulting mosaics from multiple surveys (see, e.g., \S\ref{fig:comparison}), but these methods can also be conveniently used within the $\texttt{Catalog}$ class. 

The final step in the mosaic procedure generates two TIFF color images as described in \S\ref{sec:best_low}.  Since the sensitivity of each image colored filters is survey-dependent, users who wish to extend the pipeline to mosaic images from another survey must empirically test the parameters to determine the optimal STIFF parameters that best capture the details in the telescope-specific imaging by following the guidelines in \citet{stiff}. 

\subsubsection{Catalog}

The simplest abstract class contained in the pipeline is the $\texttt{Catalog}$ class, which simply contains a list of objects in a particular catalog. While the use of such an abstract container class might seem superfluous, by using this class, the pipeline is able to cleanly separate the basic mosaic functionality for individual galaxies from the functionality required for an entire catalog. This capability enables a direct study of a single object, enables a processing of all sources listed in the derived $\texttt{Catalog}$ class, or simplifies the debugging process. By using this abstraction barrier, we ensure minimal changes in the class structure if the pipeline is modified to support imaging data from a new survey.

\section{Results}\label{results-sec}

\begin{figure}[h]
\centering
\includegraphics[width=0.4\textwidth]{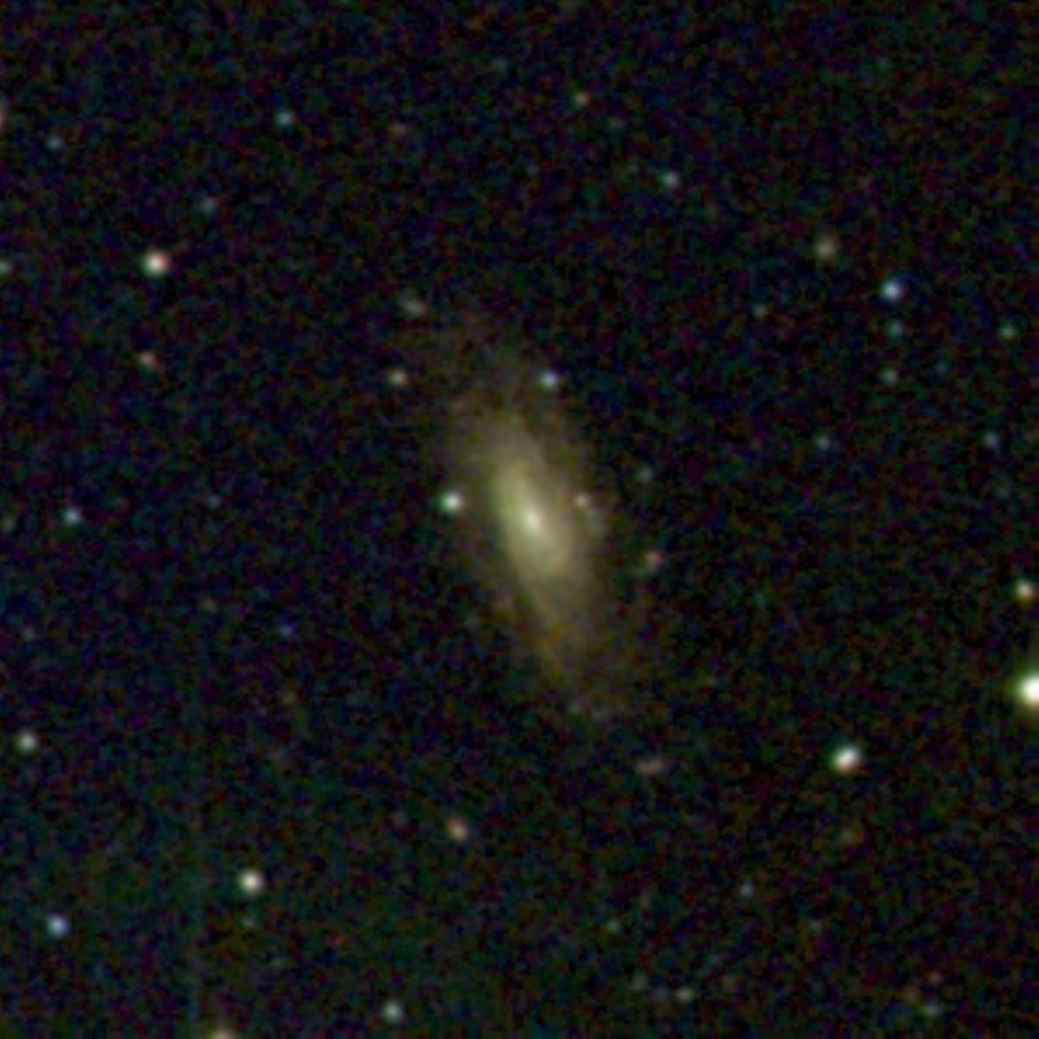}	
\includegraphics[width=0.4\textwidth]{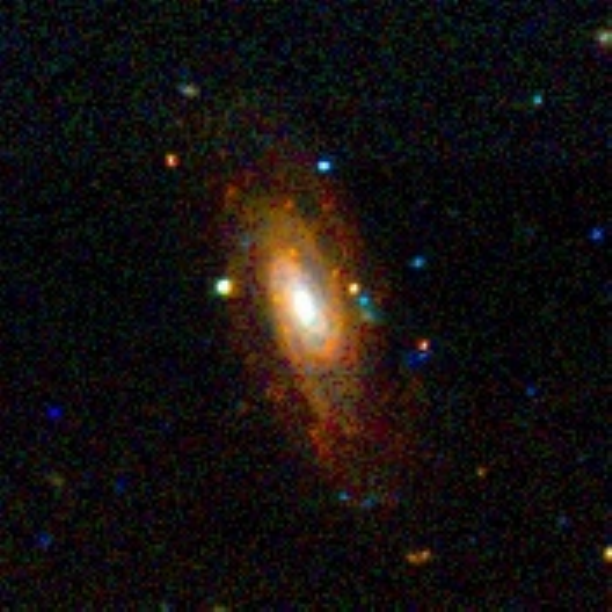}
\caption{Sample color mosaics of PGC1746 mosaic from POSS-II (top) and SDSS (bottom). The two images are of different scale.}
\label{sdss_dss_comp}
\end{figure}

\subsection{Mosaicking Results}

Of the total 23,011 galaxies described in the RC3 catalog, our automated pipeline sucessfully generated 9335 RC3 mosaics that lie within the SDSS DR10 footprint.  Of these, 4,283 RC3 galaxies were mosaicked by using an updated position that was more than one arcminute away from the originally-recorded RC3 astrometric position. On an 8-core linux server, the pipeline averaged 80 RC3 galaxy mosaics per hour for the SDSS data, which were retrieved from a remote SDSS data repository. The final data products, the five band FITS mosaic images and two color composite images for each galaxy, occupy approximately twenty GBs.

The POSS-II data has full sky coverage, thus all RC3 galaxies are covered by the POSS-II footprint. Of this full sample, the automated pipeline was successful for 99.54\% of all RC3 galaxies. For the POSS-II mosaics, 3,431 RC3 galaxies were mosaicked by using an updated position that was more than one arcminute away from the recorded RC3 astrometric position. Given the larger base images for the POSS-II survey ($\sim1.1$ GB each), it is not surprising that our pipeline was slower when creating POSS-II mosaics, averaging about fifty galaxies per hour on the same machine used for the SDSS mosaicking. The finished data products for the POSS-II sample of galaxies occupies approximately 10 GBs.

\subsection{Pipeline Performance}	

The majority of the processing time for our pipeline is in transferring the raw FITS image data from the survey data site to the processing site. However, we have explored techniques to improve the overall performance of the pipeline. First, we accelerate the mosaicking process by performing the positional update on only a single band from the bands available from a given survey. The target band is designated as the $\texttt{best\_band}$, and given the results from this single band, the other image FITs mosaics are performed only once per object. For example, for the SDSS we use r-band images, since the r-band filter since transmission has the highest quantum efficiency~\citep{edr}.

Since data transfer dominates the overall processing time, we do not employ traditional parallel programming techniques. However, since mosaic images are constructed independently, the user can employ embarrassingly parallel techniques by dividing the whole catalog into multiple smaller subcatalogs as inputs to the pipeline. Thus, even though Montage's modular design enables its performance to scale with the number of processors~\citep{montage}, our pipeline would not be accelerated by using  Message Passing Interface (MPI).  Other factors that affect the runtime include the sky coverage of a particular survey and the response speed of query from a survey archive.

One technique that could be employed to accelerate our pipeline would be local image caching, or alternatively the capability of executing our pipeline within a survey archive. We have designed our pipeline by using a class hierarchy that enables subclassing of the $\texttt{Server}$ class to specify the location of a survey's raw FITS images. However, downloading an entire survey's imaging data set for this purpose is unlikely to be beneficial (if done solely for this task), since that would likely take significantly more time than simply downloading the required input raw FITs imaged required for the mosaics when running the pipeline.

\subsection{Implementation Details}

Due to the recent rise in the popularity of the Python programming language, we developed our pipeline by using Python 2.7.6. As a result, our pipeline's dependencies are widely supported, which simplifies the extension of our work to future datasets. The majority of the mosaicking actions are done by calls to the Montage API along with the AstroPy Montage wrapper\footnote{\url{http://www.astropy.org/montage-wrapper/}} developed by~\cite{montpy}. The final color-composite images are generated by using Astromatic's STIFF v.2.4~\citep{stiff}. 

Our combination of these two programs allows us to make use of their best features. Montage excels at creating scientifically-calibrated images that retains the astrometry and photometry of input sources during the image reprojection. Our choice was also aided by the fact that the efficacy of montage was already demonstrated by the montage developers by using SDSS and POSS-II image data sets~\citep{montage}. On the other hand, STIFF provides the flexibility of adjusting a number of parameters to optimize the appearance of the final color-composite image, and also automatically estimates the upper and lower limits for the dynamic range of the final color image by using statistics derived from a pixel histogram. 

Other major steps in our pipeline that required custom code development include query construction, query result processing, source extraction on the mosaicked image, and the development of a web-accessible database to facilitate access to our data products. We use the SDSS Command line query tool written by~\cite{sqlclref}  to submit SQL queries to SkyServer and we use Astroquery as a more general archive query tool to find other RC3 galaxies that lie within a field by using the VizieR catalog database. For IRSA and most web databases, querying consists of building a URL string, submitting the query, and parsing the resulting  raw text or XML file returned by \texttt{wget}. Source extraction was performed by using SExtractor v.2.19.5~\citep{sextractor} with standard processing parameters. Our web-accessible database was created by using the Python sqlite3 module, and the web search interface was written in PHP with interacting HTML elements.

\subsection{Science Quality of Output Mosaic\label{preserved}}
 As described by  ~\cite{montage}, the Montage algorithms used for reprojection, background rectification, and coaddition preserve the astrometry and photometry of the input sources in the resulting mosaics. Montage has been third-party validated by many survey projects, including Spitzer Wide Area Infrared Experiment (SWIRE) and COSMOS Cosmic Evolution Survey, which used Montage to generate science-grade data products. As a separate, additional test, we match together the relevant quantities, as measured by SExtractor, in the input and output images. We chose to only conduct our analysis on the SDSS survey and not the DSS since, as discussed in Sec. \ref{POSSII}, we are only cropping the digitized plate images and not mosaicking these images, so the astrometry and photometry must be preserved along with FITS  header for all DSS data.
\subsubsection{Astrometric Preservation\label{astrometry}}

\begin{figure}[h]
\includegraphics[width=0.5\textwidth]{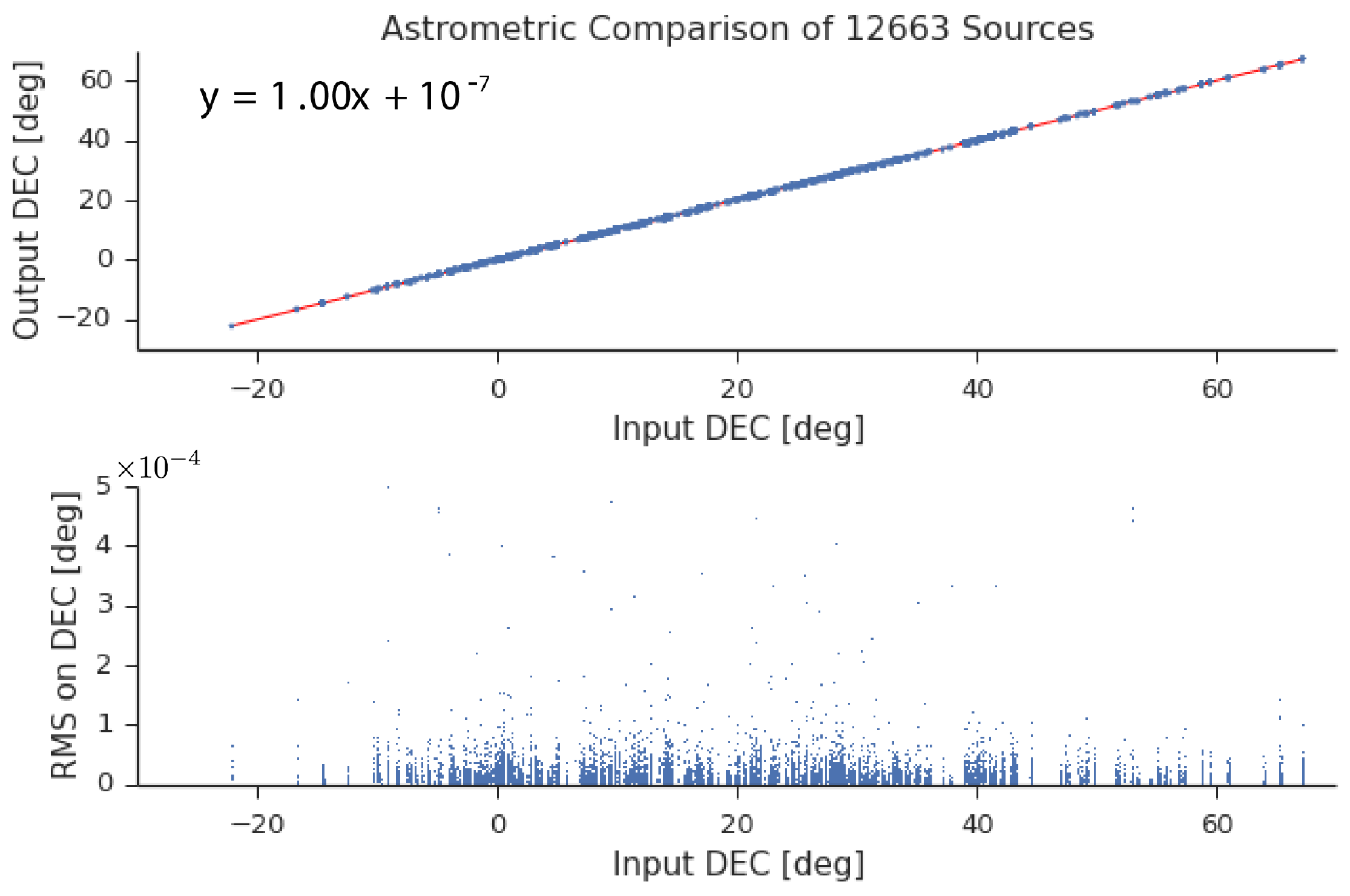}
\caption{Two plots demonstrating the preservation of astrometric calibration between input images and output mosaics for 12,663 RC3 sources detected in SDSS. (Top) The direct relationship between input and output declination coordinate values, measured in degrees. (Bottom) The residual RMS difference between input and output declination values as a function of input declination, showing no systematic effects across the sky.}
\label{ra}
\end{figure}
To test whether the pipeline introduces errors in the astrometry, we mosaiced a randomly-selected set of 500 RC3 galaxies with a variety of sizes and uniformly distributed around the sky and match them with the detected sources in the input fields found by SExtractor.  Figure~\ref{ra} shows the one-to-one correlation between the input and output declination. A similar trend was observed between the input and output right ascensions, which yielded a linear regression with a slope of one and an offset of $3.13\times10^{-7}$ degree. Further, we find that there is no systematic offset in the RMS of the input and output coordinates down to an order of  $10^{-5}$ degrees. An offset at this level is negligible and probably due to floating point arithmetic error.

\subsubsection{Photometric Calibration\label{photo}}

We use Montage's default values to handle image coaddition, which uses the area-weighted average of the input fluxes to compute the output pixel intensity.  Montage does not perform PSF matching across different imaging fields since that would require telescope-specific detail about the survey, as the PSF can vary across large images due to the camera optics. Since we are using calibrated SDSS data and mosaicking neighboring fields, any PSF variability will be minimal and will not affect our mosaics since the RC3 galaxies are all relatively large compared to the survey PSFs. Thus, we do not expect any changes to the photometric calibration of the images upon which our pipeline operates.

\begin{figure}[h]
\includegraphics[width=0.5\textwidth]{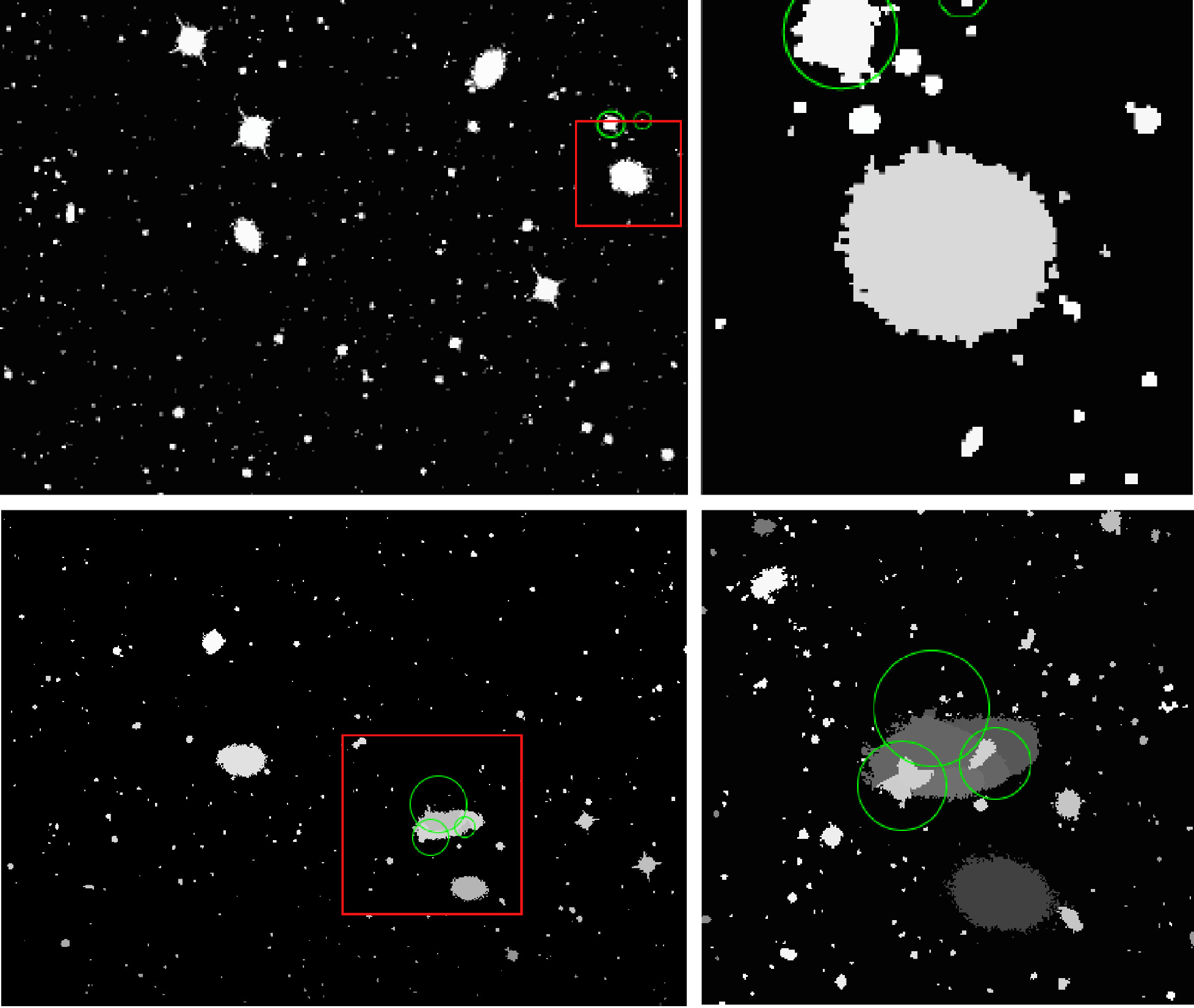}
\caption{A visual demonstration of two issues affecting source photometry between input and output images for two different RC3 galaxy fields. The left-hand column shows the SExtractor \textit{check-image} for the SDSS input image field for PGC 5095 (top) and PGC1921 (bottom). In both of these images, the red bounding box highlights the check image (shown on the right) for the output image mosaic, with sources detected by SExtractor that have significantly different photometric measurements between the input and output images circled in green. The PGC5095 field demonstrates how an output image mosaic can crop sources, while the PGC1921 field demonstrates how deblending and background calculations can be affected on an output image (which in this case is considerably smaller than the input image field).}
\label{outlier_rejection}
\end{figure}

We explore this hypothesis by performing a similar analysis as described in \S\ref{astrometry}, except in this case we compare the SExtractor-computed values for corrected isophotal magnitudes ($\texttt{MAG ISOCORR}$). As shown in Figure~\ref{outlier_rejection}, photometric measurements are affected by several different issues, including source blending and background estimation. For example, when a source is near the edge of an output mosaic image, object pixels can be missed or the local background can be misestimated, both of which will effect a photometric measurement for that source. For a direct comparison of input and output photometric measurements, we therefore remove all sources that lies near the boundary of an image field.

In addition, by construction, most source detection software, including SExtractor, have trouble detecting and deblending large, extended sources such as RC3 galaxies. One possible reason for this is that the input image field can be (potentially much) larger than the cropped output image field that our pipeline saves for each RC3 galaxy. Thus, even if we use the same deblending threshold for SExtractor, the different images will have different photometric measurements since the background level has changed, which affect both source detection and photometry. This will result in large RMS deviations between the input and output image photometry; we therefore automatically reject any source that clearly has wrongly deblended sources that are close to an RC3 galaxy.
	
With filters in place to remove these \textit{catastrophic} outliers, we performed a photometric analysis of a number of selected fields at various right ascension and declinations. In total these fields contained 12,663 detected and photometered sources over 389 different SDSS fields. Of these sources, 427 were removed as being edge-affected and 94 were removed for being too close to a target RC3 galaxy. Our confidence in these two filters is reinforced since after the edge-affected outliers were removed, the number of outlier at bright magnitudes (-6$\sim$-10) with RMS from 0.5$\sim$1.5 decreased. After removing the sources too close to an RC3 galaxy, the number of outliers at faint magnitudes (-4$\sim$0) with RMS from 1.5$\sim$3.0 decreased, while still preserving correctly-deblended RC3 sources.

\begin{figure}[h]
\includegraphics[width=0.5\textwidth]{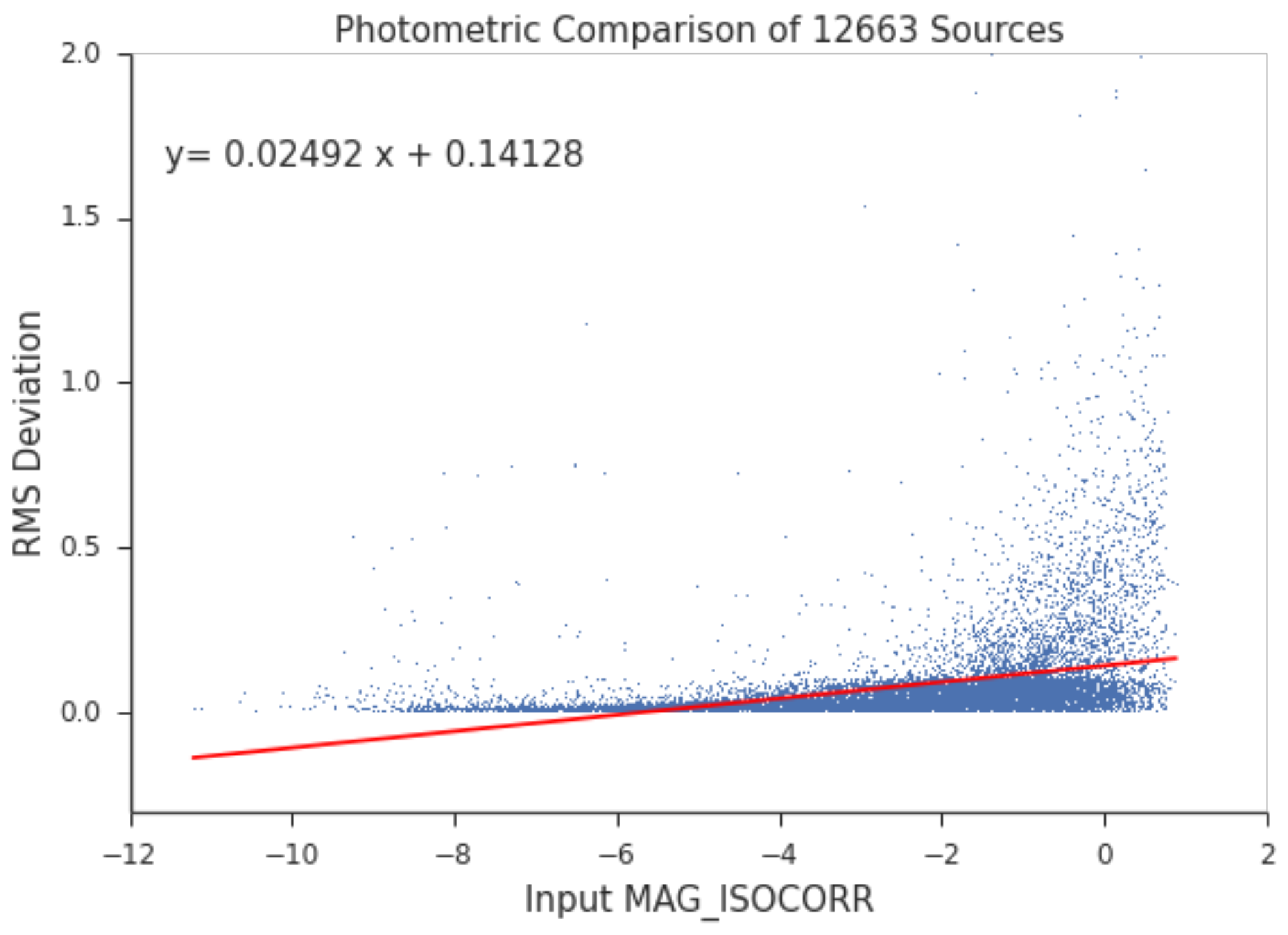}
\caption{The RMS magnitude deviation as a function of raw input magnitude demonstrating the relationship between input and output photometry, which are predominantly consistent across the measured magnitude range.}
\label{photo_preserved}
\end{figure}

As shown in Figure~\ref{photo_preserved}, the RMS spread of the magnitude difference between the photometry on the input and output images for these selected fields is remarkably consistent. Since the two images are generally of different size and the objects have shifted between the two images, we do not expect a perfect one-to-one correspondence, especially with this naive photometric comparison (in contrast, for example, to the full SDSS photometric pipeline). While there are still some sources with larger magnitude differences than we might naively expect, the fact that this comparison is simply a direct comparison between two SExtractor runs means the two measurements do not necessarily agree on pixel deblending, which pixels belong to which source, nor on the overall background in each image. 

\begin{figure}[h]
\includegraphics[width=0.5\textwidth]{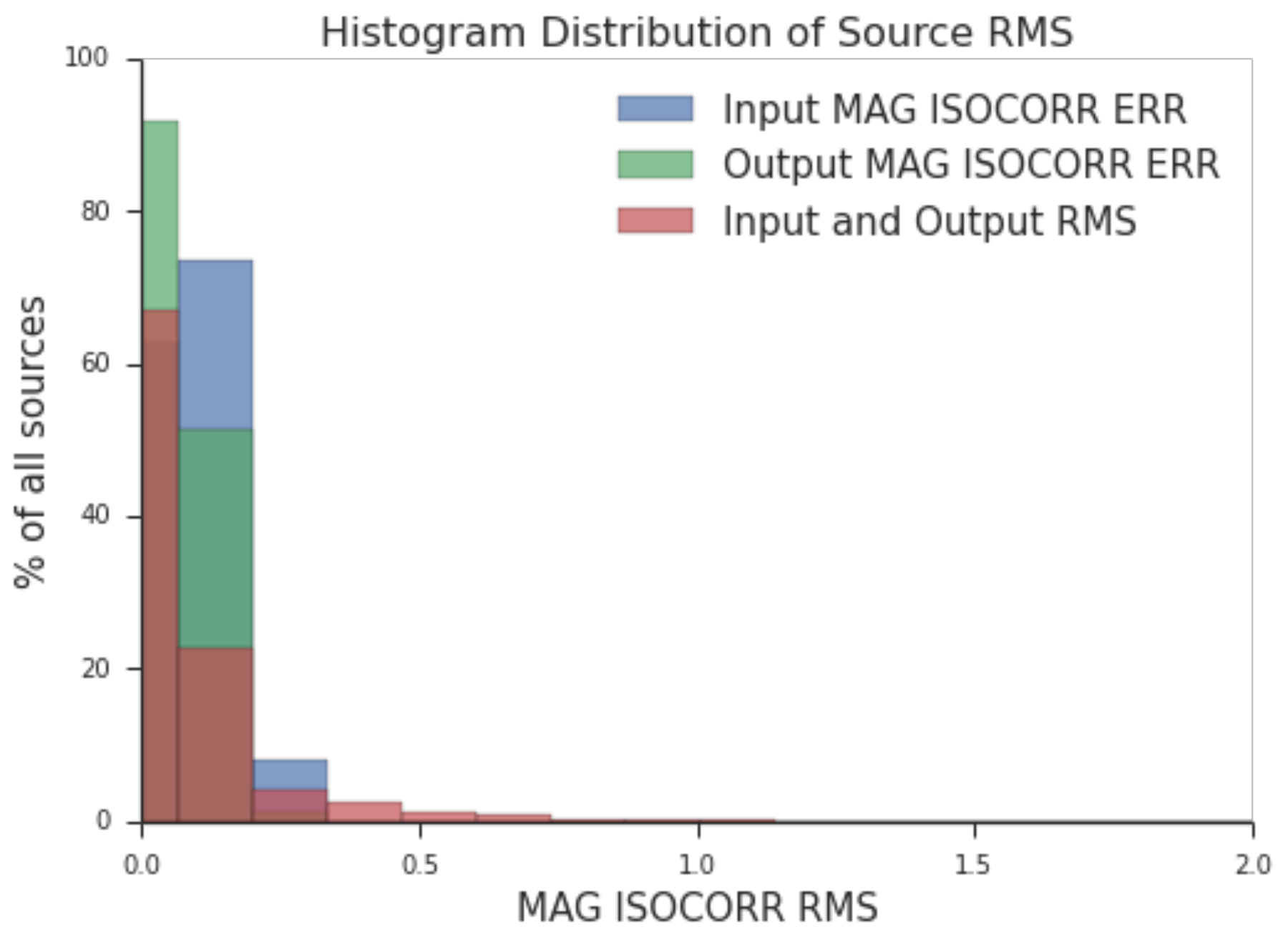}
\caption{A histogram of the magnitude RMS deviations between input and output images as computed by SExtractor. The bulk of the photometric differences can be assigned to the differences in the SExtractor photometric measurement between the two images for RMS $<0.25$, while the larger outliers correspond to different sized images, pixel assignment conflicts, or deblending issues.}
\label{histo_rms}
\end{figure}

We also note that the average RMS magnitude difference for the whole sample  ($\sim$ 0.086) is comparable to the mean error on the \texttt{MAG ISOCORR} measurement returned by SExtractor (0.088 on the input and 0.058 on the output). We compare these statistical samples in Figure~\ref{histo_rms}, where we clearly see the general agreement between input and output image photometry, which is as expected since montage preserves the total intensities of all the pixels between the input and output images. Thus, the deviation for sources with RMS$<$0.25 can largely be assigned to errors inherent in the photometry computed by SExtractor, while the less numerous, larger amplitude outliers are the result of varying image background levels, differences in deblending, or differences in pixels being assigned to objects for source photometry.

 \section{Conclusion}\label{conc-sec}
 
By design, wide-area sky surveys are used to answer fundamental questions regarding the formation and evolution of large-scale structure and of the cosmological history of the universe. But these data can be used for many other purposes. The mosaicking pipeline described in this paper, for example, provides a convenient way to generate mosaic images for specific sources from an archived set of sky survey images.  Furthermore, this pipeline can be easily adapted to work with future data to create scientifically-calibrated FITS mosaics as well as color-composite images. The source code and documentation for the pipeline described in this paper can be found in the project repository\footnote{\url{http://github.com/ProfessorBrunner/rc3-pipeline/}}. In addition, we have provided documentation on GitHub that will guide other investigators to adapt the pipeline for alternative imaging data sets.

To ensure that the generated mosaic images are centered on the target source, we implemented an algorithm that automatically determines the correct source astrometry, updates the source catalog appropriately, and generates mosaic images centered on the newly updated coordinate location. Finally, to demonstrate the efficacy of this new pipeline, we generated FITS image mosaics and color-composite images of galaxies in the RC3 catalog by using the SDSS and POSS-II data, along with the most-up-to date set of positional values for all the RC3 galaxies. All of these data products are publicly released and accessible via a searchable web form on the Laboratory for Cosmological Data Mining website. \footnote{\url{http://lcdm.astro.illinois.edu/data/rc3/search.html}}

By developing this new pipeline, we can generate image mosaics using newly-obtained data, which will enable a more complete sky coverage for a given source catalog or the construction of potentially higher resolution images or images in other wavelengths. In addition, the pipeline simplifies the extension of this work to either user-defined catalogs or to other published catalogs, such as the Messier Catalog or the New General Catalog. Furthermore, a specific scientific inquiry may require the construction of a user-defined catalog  by imposing selection criteria to study certain types of objects. To accomplish this task, a user simply needs to generate a text file containing source positions, source radii, and unique identifier for each source, which can subsequently be used as input to the pipeline. 

Other potential uses of the pipeline includes masking large RC3 galaxies to prevent CCD saturation, selection of spectroscopic targets, and generating a collection of  multi-band color images on the catalog sources. Multi-band images of the same area of the sky are useful for extracting a wealth of science information about a particular source. The scientific value of multi-band images are evident in areas such as the Stripe 82, which many  surveys have chosen to maximize their overlapping area with in order to obtain such a collection of multi-band images on the targeted sources. The generated FITS mosaic images can also be used as inputs to existing tools such as Astrometry.net~\citep{astrometry.net} or SExtractor~\citep{sextractor} for subsequent processing or to tools like ds9~\citep{ds9}, Alladin~\citep{aladin}, or APLpy~\citep{aplpy} for scientific visualization.
\section*{Acknowledgements}
\footnotesize

The work was supported by the Google Summer of Code Program. RJB would also like to acknowledge support from the National Science Foundation Grant No. AST-1313415 and support as a center associate at the Center for Advanced Study at the University of Illinois. We thank Harold G. Corwin Jr. for helpful discussion that helped this work. 

This research made use of Montage, funded by the National Aeronautics and Space Administration's Earth Science Technology Office, Computation Technologies Project, under Cooperative Agreement Number NCC5-626 between NASA and the California Institute of Technology. Montage is maintained by the NASA/IPAC Infrared Science Archive. This research made use of Astropy, a community-developed core Python package for Astronomy.

Funding for SDSS-III has been provided by the Alfred P. Sloan Foundation, the Participating Institutions, the National Science Foundation, and the U.S. Department of Energy Office of Science. The SDSS-III web site is http://www.sdss3.org/. SDSS-III is managed by the Astrophysical Research Consortium for the Participating Institutions of the SDSS-III Collaboration including the University of Arizona, the Brazilian Participation Group, Brookhaven National Laboratory, University of Cambridge, Carnegie Mellon University, University of Florida, the French Participation Group, the German Participation Group, Harvard University, the Instituto de Astrofisica de Canarias, the Michigan State/Notre Dame/JINA Participation Group, Johns Hopkins University, Lawrence Berkeley National Laboratory, Max Planck Institute for Astrophysics, Max Planck Institute for Extraterrestrial Physics, New Mexico State University, New York University, Ohio State University, Pennsylvania State University, University of Portsmouth, Princeton University, the Spanish Participation Group, University of Tokyo, University of Utah, Vanderbilt University, University of Virginia, University of Washington, and Yale University.

\bibliography{ms}
\bibliographystyle{apj}
\end{document}